\documentclass[12pt]{iopart}
\usepackage{graphics}
\usepackage{epsfig}
     
     \newcommand{\pathnow}{}
\def\lessim{\lower.5ex\hbox{$\; \buildrel < \over \sim \;$}}
\def\gtrsim{\lower.5ex\hbox{$\; \buildrel > \over \sim \;$}}
\newcommand{\nc}{\newcommand}
\nc{\ds}{\displaystyle}        \nc{\ts}{\textstyle}
\nc{\rf}[1]{Fig.\,\ref{#1}}    \nc{\rt}[1]{table\,\ref{#1}}
\nc{\req}[1]{Eq.\,(\ref{#1})}  \nc{\eps}{\varepsilon}
\nc{\beq}{\begin{equation}}     \nc{\beql}[1]{\begin{equation}\label{#1}}
\nc{\eeq}{\end{equation}}        
\nc{\beqa}{\begin{eqnarray}}   \nc{\eeqa}{\end{eqnarray}}       
\nc{\bfi}{\begin{figure}}       \nc{\efi}{\end{figure}}

\begin{document}
\title{Strangeness and threshold of phase changes}
\author{Johann Rafelski$^1$, Inga Kuznetsova$^1$, and Jean Letessier$^{1,2}$}%
\address{$^1$ Department of Physics, University of Arizona, Tucson, AZ 85721\\
$^2$ 
LPTHE, Universit\'e Paris 7, 2 place Jussieu, F--75251 Cedex 05}

\ead{rafelski@physics.arizona.edu}

\begin{abstract}
{We explore entropy and strangeness as signature of QGP for top AGS and 
the energy scan at SPS. We find that the hadronization dynamics changes between
20 and 30 $A$ GeV projectile energy. The high energy results are consistent with QGP.
} 
\end{abstract}
\pacs{24.10.Pa, 25.75.-q, 13.60.Rj, 12.38.Mh}
\section{Introduction} \label{intror}
It is commonly accepted that, at RHIC, we have produced  the quark-gluon
plasma (QGP) in laboratory; the question is
if this new state of matter is already present in the  SPS energy range as has been 
considered likely in the CERN press release of February 2000? We present here a brief discussion
of our findings as described in full length elsewhere~\cite{Letessier:2005qe,Kuznetsova:2006bh}.  
Here, we look at  the abundant  production of  strange flavored hadrons 
in the relativistic heavy ion  collisions at  top AGS and all SPS energies, addressing the 
NA49 energy scan results \cite{NA49Gaz}.  

Our  data analysis employs the  statistical hadronization
model (SHM); we assume that the strong interactions  saturate the   quantum particle production
matrix elements. Therefore,  the  yield  of   particles  is controlled dominantly by the magnitude of the
accessible phase space. The SHM contains little if any 
information about the nature of interactions, and thus, it embodies the 
objective of reaching simplicity in many body dynamics, allowing to 
identify the properties of the dense and hot primary matter formed in 
heavy ion collisions.  Interpretation of experimental data is arrived at with 
 the SHARE suite of programs \cite{Torrieri:2004zz}.

\section{Strangeness and Entropy}\label{sssec}
The total final state hadron multiplicity is a  measure of the entropy $S$ produced. 
In the  QGP, the entropy  production occurs predominantly 
early on in the collision, 
once a quasi-thermal exponential energy  distribution of partons 
has been formed, the entropy production has been mostly completed, even 
if the chemical yield equilibrium is not achieved yet.
Since the kinetic processes leading to strangeness production are 
slower than the parton equilibration process, we are
rather certain that the production of entropy occurs mainly prior to strangeness 
production.  Even though the degree of chemical  equilibration   of
gluons in early stages  (which dominate strangeness production) are uncertain, 
study of kinetic strangeness production show that the controlling 
quantity   is the entropy contents. For this reason,  the observable 
`strangeness pairs per entropy' $N_s/S$ (also colloquially referred to as $s/S$)
emerges as diagnostic tool, also since in essence both $s$ and $S$ are conserved
in the process of hadronization.

The phase space density is in general different in any 
two matter phases. Thus when transformation of one phase `Q' (for QGP) 
into the other occurs rapidly,  given chemical equilibrium in the
decaying phase, in general the final state is out-of chemical equilibrium. 
Especially, when  hadrons  are produced in a recombinant model, 
in order to preserve entropy, there must be  a jump in the phase
 space occupancy parameters $\gamma_i^Q<\gamma_i, i=q,s$.
The  superscript  $Q$ indicates that we refer  to the QGP phase; variables
in the hadron phase will be stated without an superscript.  

This jump replaces the increase in volume found 
in a slow transformation involving  re-equilibration.
In order to preserve entropy in sudden hadronization of supercooled QGP  at  $T\simeq 140$ MeV,
we  must have  for the light quark `q' occupancy $\gamma_q^{\rm cr}=e^{m_\pi/T}$.
The value $\gamma_q^{\rm cr}$  is where the pion gas  condenses. The required value of  $\gamma_q$ is
decreasing with increasing temperature
and is crossing  $\gamma_q=1$  near $T\simeq 180$ MeV.
Thus, in fast hadronization of the QGP phase (without an increase in volume, i.e., mixed phase), 
we expect that, the value of $\gamma_q$ governing hadron yields must  
be greater than unity for every value of $T$ considered in previous studies of 
the hadronization process, with the relation being approximately as
$ 
\gamma_q\simeq1.6-0.015 (T-140)\,[\mathrm{MeV}]
$. 

The ratio $s/S$ up to a structural numerical factor compares the degeneracy
  of strangeness to the overall QGP effective degeneracy.  
At sufficiently high temperature, the entropy density $S/V$ in QGP is that of 
(nearly) ideal quark-gluon gas:
\begin{equation}\label{S1}
{S\over V}={4\pi^2\over 90} g_{\rm eff}^Q(T)T^3={\rm Const.},
\end{equation}
For  an   equilibrated 
QGP phase with perturbative properties:
\beql{sdivS}
{s \over S}\equiv\frac{\rho_{\rm s}}{S/V}   \simeq
\frac{ (\gamma_s^Q(t) g_s^Q/\pi^2) T^3 0.5\,x^2K_2(x)}
  {g_{\rm eff}^Q\,4\pi^2/ 90\, T^3}
=\frac{\gamma_s^Q g_s^Q}{g_{\rm eff}^Q} 0.23 [0.5 x^2K_2(x)]\,.
\eeq
For early times, when $x=m_s/T(t)$ is relatively small, assuming the equilibrium value ($\gamma_s^Q=1$),
we can find  $s/S\simeq 0.045$. However, at high temperature strangeness is not yet equilibrated
chemically and in general the value in QGP at hadronization is expected in the range $0.03<s/S<0.04$.
When and if strangeness is not equilibrated in the QGP source, we in effect can determine the 
value $\gamma_s^Q$ by comparing to the above expectations. 

\begin{figure}
\centering
\includegraphics[width=8.5cm,height=8.cm]{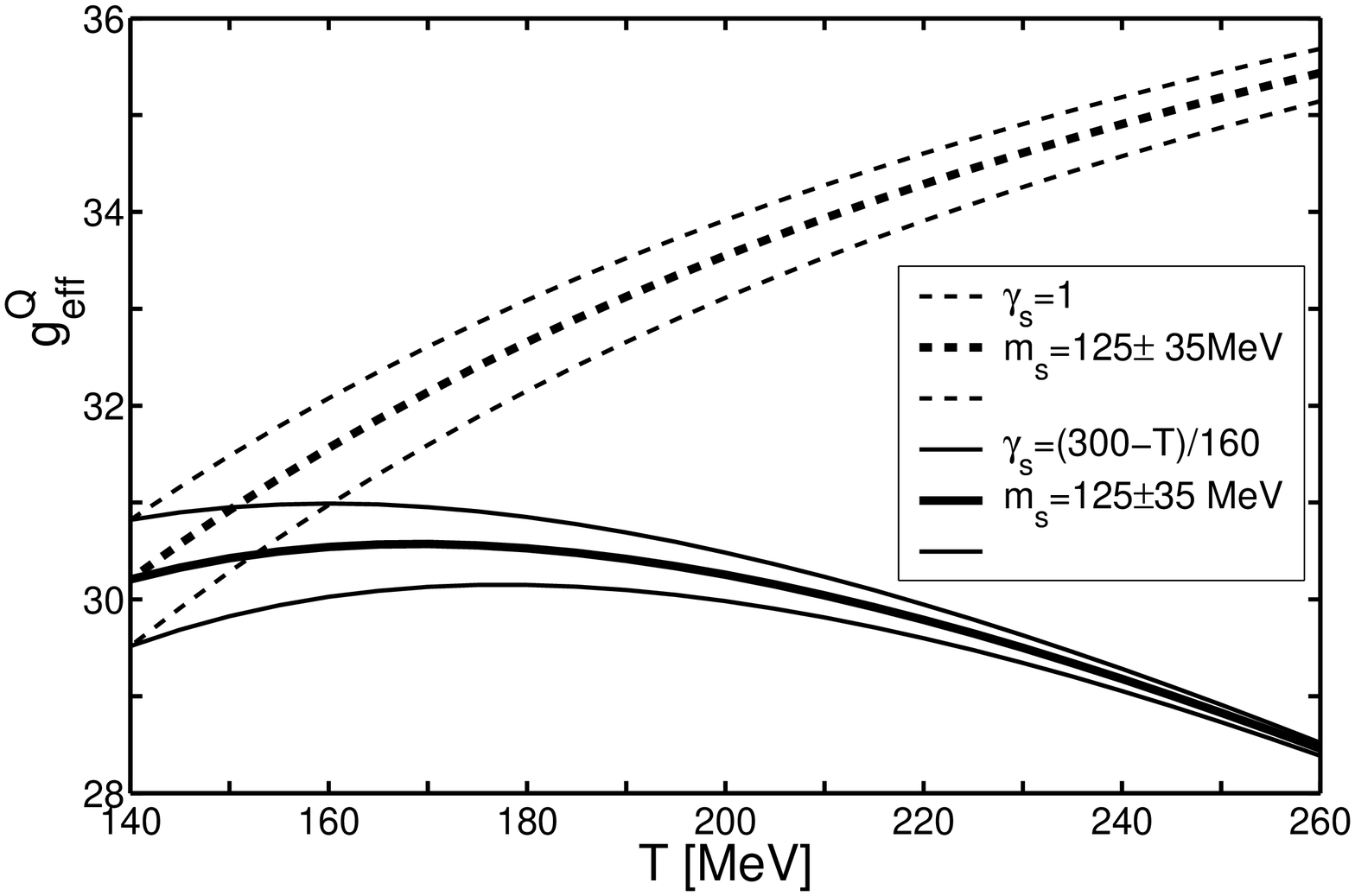}
\caption{ 
\small{
The Stefan-Boltzmann degrees of freedom $g_{\rm eff}$  as function of temperature $T$ 
for  $m_s=125 $\,MeV (central thick  lines),  thin upper and lower 
lines for $m_s=90$and $160$ \,MeV respectively. The  dashed lines are for  chemically equilibrated
$u,\ d,\ s$ and $G$ QGP plasma, with perturbative corrections for degeneracy. The solid lines are
for QGP in which strangeness contents is increasing as 
temperature is decreasing, approaching chemical equilibrium. }} \label{geff}
\end{figure}

In order to arrive at the above estimate, we needed to use 
the   number of degrees of freedom in chemically equilibrated QGP, which is 
shown by dashed lines in \rf{geff}. Solid lines allow for the effect of approach 
to chemical equilibrium of strangeness, assuming:
$
\gamma_s^Q\simeq (300-T)/{160}\,\mathrm{MeV} 
$. 
Most of temperature dependent corrections 
cancel, and one finds in the latter case  that it is possible 
to use a nearly  $T$ independent value seen for for $T<260$ MeV in \rf{geff},  $g_{\rm eff}^Q\simeq 30$ 
near to QGP breakup condition, which value is decreasing  to  $g_{\rm eff}^Q\simeq 28$ for a hot  
QGP~\cite{Kuznetsova:2006bh}. 

To quantify the strangeness enhancement signature due to deconfinement we compare in \rf{sST} the value of $s/S$ in 
chemically equilibrated hadron matter with that of chemically equilibrated QGP, see \rf{sST},
as function of chemical freeze-out temperature $T$. This figure quantifies the specific strangeness
enhancement of the QGP phase. This enhancement implies in  fast hadronization  that a chemical nonequilibrium 
must arise among (strange) hadrons formed. The high density of strangeness present at QGP hadronization 
can  therefore lead to a considerable enhancement of  the  yields of 
multi-strange antibaryons, and $\phi$~\cite{Rafelski:1982ii}. 
Strangeness enhancement is best expressed by the magnitude of 
$\gamma_s/\gamma_q$ after hadronization, which we evaluate  conserving strangeness and entropy.
at hadronization, beginning with a   QGP phase which is nearly chemically equilibrated at the
point of hadronization: 
\beql{sSapp}
s/S=f(\gamma_s,\gamma_q,T)\simeq  (\gamma_s/\gamma_q)^{5/6}(0.026\pm0.01), \quad 
 T\in (140,180) {\rm MeV}.\eeq
\begin{figure}
\centering
\includegraphics[width=9cm,height=8cm]{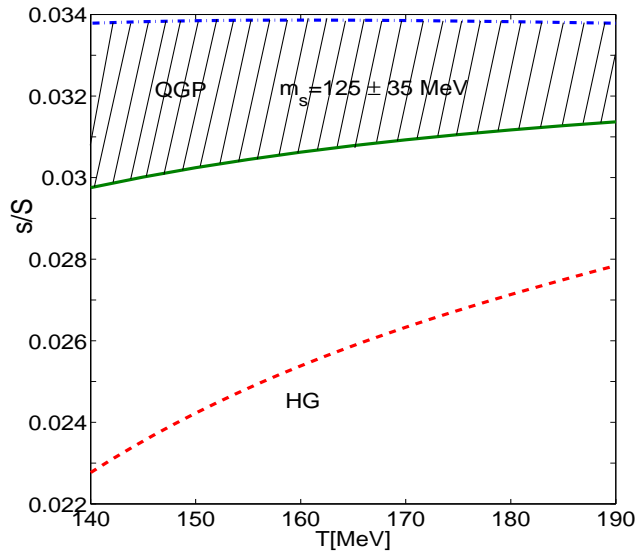}
\caption{Strangeness to entropy ratio $s/S(T; \mu_B=0,\mu_S=0)$
for the chemically equilibrated QGP (green, solid line for $m_s=160$ MeV, 
blue dash-dot line for $m_s=90$ MeV);
 and for chemically equilibrated HG (red, dashed).}\label{sST}
\end{figure}

Only when one considers the entropy 
content as measured by hadron yields, the understanding of the strangeness 
production in QGP is reliable. If instead  the production of
strangeness  is studied at some presumed `hydro'-inspired 
space-time profile of temperature $T$,  results are meaningful  only if  
the QGP entropy $S$  content has been directly related  to
an observed hadron multiplicity. Namely, if  the initial temperature 
is modeled to be  20\% below value needed for the observed entropy, 
the initial entropy content is cut in half. In the   QGP this means that there are half
as many gluons, and the rate of strangeness production by gluon fusion  is cut down by 
a factor 4, which in turn implies that strangeness  would not equilibrate in QGP.  

\section{Energy Scan at CERN-SPS}\label{NA49fit}
We analyze the particle  production obtained in the 
 energy scan of the NA49 experiment 
at CERN-SPS~\cite{NA49Gaz}, and also include the results of our earlier 
analysis of the top AGS data~\cite{Letessier:2004cs}. The outcome
of the fit procedure is stated in the top section  of table \ref{AGSPS}.
The $\lambda_s$  values,  marked with
an asterix $^*$ in  table \ref{AGSPS}, are result of a strangeness conservation constraint, 
which, however, is not chosen to be zero, but as shown in table: 
since strangeness conservation constraint involves 
several particle yields it is inappropriate to insist on $s-\bar s =0$, since this 
correlates the errors of the input data which are experimentally not correlated. Our procedure was to
fit first without strangeness conservation, and once we see the strangeness asymmetry
to fix it at the best value shown in table so that the resulting procedure of fixing $\lambda_s$ is
the same as used by other groups, but that there is no unnecessary error constraint
introduced among strange hadrons.

\begin{table}[ht]
\centering
\caption{For each projectile energy $E$ [$A$\,GeV] for AGS and SPS energy range, we  present in the header  
$\sqrt{s_{\rm NN}}$, the invariant center of momentum
 energy per nucleon pair,   $y_ {\rm CM} $  the center of
momentum rapidity. This is followed
by statistical parameters $T, \lambda_i, \gamma_i$  obtained in the fit, the strangeness asymmetry required,  
and we present the resulting 
 chemical potentials $\mu_{\rm B}, \mu_{\rm S}$,  the reaction volume $V$ and 
the centrality of the reaction considered. This is  followed    first by input and than by output 
total hadron multiplicity $N_{4\pi}$.\\}\label{AGSPS}
\scriptsize

\begin{tabular}{|c| c | c c c c c |  }
\hline
E[$A$GeV]                      & 11.6          & 20          & 30              & 40         & 80            & 158  \\
$\sqrt{s_{\rm NN}}$  [GeV]     &4.84           &6.26         &7.61             &8.76          &12.32          &17.27  \\
$y_{\rm CM}$                   &1.6  &1.88 &2.08 &2.22 & 2.57& 2.91 \\
\hline
$T$ [MeV]                   &157.8$\pm$0.7  &153.4$\pm$1.6  &123.5$\pm$3    &129.5$\pm$3.4   &136.4$\pm$0.1 &136.4$\pm$0.1       \\
$\lambda_q$                 &5.23$\pm$0.07  &3.49$\pm$0.08  &2.82$\pm$0.08  &2.42$\pm$0.10   &1.94$\pm$0.01 & 1.74$\pm$0.02     \\
$\gamma_q$                 &0.335$\pm$0.006  &0.48$\pm$0.05  &1.66$\pm$0.10  &1.64$\pm$0.04  &1.64$\pm$0.01 &1.64$\pm$0.001\\
$\gamma_s$                  &0.190$\pm$0.009 &0.38$\pm$0.05  &1.84$\pm$0.32  &1.54$\pm$0.15  &1.54$\pm$0.05 & 1.61$\pm$0.02     \\
$\lambda_{I3}$             &0.877$\pm$0.116  &0.863$\pm$0.08 &0.939$\pm$0.023&0.951$\pm$0.008&0.973$\pm$0.002& 0.975$\pm$0.004     \\
\hline
$\lambda_s$                 &1.657$^*$      &1.41$^*$       &1.36$^*$       &1.30$^*$        &1.22$^*$      & 1.16$^*$   \\
$s-\bar s/s+\bar s$    & 0    &-0.092&-0.085 &-0.056&-0.029&-0.062  \\
\hline
$\mu_{\rm B}$ [MeV]               &783      &576            &384            &344             &271           & 227   \\
$\mu_{\rm S}$ [MeV]               &188      &139            & 90.4          &80.8            &63.1          & 55.9  \\
\hline
\hline
$V {\rm [fm}^3]$            &3596$\pm$331   &4519$\pm$261   &1894$\pm$409   &1879$\pm$183    &2102$\pm$53   & 3004$\pm $1  \\
$N_{4\pi}$ centrality          &most central   &  7\%          &  7\%          &  7\%           &  7\%           &  5\%\\
\hline
$R=p/\pi^+$,  $N_W $
                        &$R=1.23 \pm 0.13$  &349$\pm$6      &349$\pm$6      &349$\pm$6      &349$\pm$6      &362$\pm$6           \\
$Q/b $                      &0.39$\pm$0.02  &0.394$\pm$0.02 &0.394$\pm$0.02 &0.394$\pm$0.02 &0.394$\pm$0.02 &0.39$\pm$0.02   \\
$\pi^+$                     &133.7$\pm$9.9  &184.5$\pm$13.6 &239$\pm$17.7   &293$\pm$18     &446$\pm$27     &619$\pm$48       \\
$R=\pi^-$/$\pi^+$, $\pi^-$
                        & $R=1.23 \pm 0.07$ &217.5$\pm$15.6   &275$\pm$19.7 &322$\pm$19     & 474$\pm$28     &639$\pm$48       \\
$R={\rm K}^+\!/{\rm K}^-$,${\rm K}^+$   
                            &$R=5.23\pm0.5$  &40$\pm$2.8     &55.3$\pm$4.4   &59.1$\pm$4.9   &76.9$\pm$6     &103$\pm$10      \\
${\rm K}^-$                 &3.76$\pm$0.47  &10.4$\pm$0.62  &16.1$\pm$1     &19.2$\pm$1.5   &32.4$\pm$2.2   &51.9$\pm$4.9      \\
$R=\phi/{\rm K}^+$, $\phi $ 
                         &$R=0.025\pm 0.006$&1.91$\pm$0.45  &1.65$\pm$0.5   &2.5$\pm$0.25   &4.58$\pm$0.2   & 7.6$\pm$1.1      \\
$\Lambda$                   &18.1$\pm$1.9   &28$\pm$1.5     &41.9$\pm$6.1   &43.0$\pm$5.3   &44.7$\pm$6.0   &44.9$\pm$8.9       \\
$\overline\Lambda$          &0.017$\pm$0.005&0.16$\pm$0.03  &0.50$\pm$0.04  &0.66$\pm$0.1   &2.02$\pm$0.45  &3.68$\pm$0.55    \\
$\Xi^-$                     &               & 1.5$\pm$0.13  & 2.48$\pm$0.19 &2.41$\pm$0.39  &3.8$\pm$0.260  & 4.5$\pm$0.20      \\
$\overline\Xi^+$            &               &               &0.12$\pm$0.06  & 0.13$\pm$0.04 &0.58 $\pm$0.13 & 0.83$\pm$0.04      \\
$\Omega+\overline\Omega$ //  ${\rm K}_{\rm S}$&         &               &               &0.14$\pm$0.07   &              &  81$\pm$4  \\
 \hline\hline
$b\equiv B-\overline B$    & 375.6&347.9 &349.2 &349.9 &350.3 &362.0 \\
$\pi^+$                    & 135.2&181.5 &238.7 &290.0 &424.5 &585.2 \\
$\pi^-$                    & 162.1&218.9 &278.1 &326.0 &461.3 &643.9 \\
${\rm K}^+$                & 17.2 &39.4  &55.2  &56.7  &77.1  &109.7 \\
${\rm K}^-$                & 3.58 &10.4  &15.7  &19.6  &35.1  &54.1  \\
${\rm K}_{\rm S}$          & 10.7 &25.5  &35.5  &37.9  &55.1  &80.2  \\
$\phi $                    & 0.46 &1.86  &2.28  &2.57  &4.63  &7.25  \\
$p $                       &174.6 &161.6 &166.2 &138.8 &138.8 &144.3 \\
$\bar p$                   &0.021 &0.213 &0.68  &0.76  &2.78  &5.46  \\
$\Lambda$                  & 18.2 &29.7  &39.4  &34.9  &42.2  &48.3  \\
$\overline\Lambda$         &0.016 &0.16  &0.51  &0.63  &2.06  &4.03  \\
$\Xi^-$                    & 0.47 &1.37  & 2.44 &2.43  &3.56  &4.49  \\
$\overline\Xi^+$           &0.0026&0.027 &0.089 &0.143  &0.42  &0.82  \\
$\Omega$                   & 0.013&0.068 &0.14  &0.144 & 0.27 &0.38  \\
$\overline\Omega$          &0.0008&0.0086&0.022 &0.030 & 0.083&0.16  \\
${\rm K}^0(892)  $         & 5.42 &13.7 &11.03  &12.4  &18.7  &26.6 \\
$\Delta^{0} $              & 38.7 &33.43 & 25.02 &26.6  &27.2  &28.2  \\
$\Delta^{++} $             & 30.6 &25.62 & 22.22 &24.2  &25.9  &26.9  \\
$\Lambda(1520)$            & 1.36 &2.06  & 1.73 &1.96  &2.62  &2.99  \\
$\Sigma^-(1385)$           & 2.51 &3.99  & 4.08 &4.26  &5.24  &5.98  \\
$\Xi^0(1530) $             & 0.16 & 0.44 & 0.69 &0.73  &1.14  &1.44  \\
$\eta $                    & 8.70 &16.7  & 19.9 &24.1  &38.0  &55.2  \\
$\eta' $                   & 0.44 &1.14  & 1.10 &1.41  &2.52  &3.76  \\
$\rho^0 $                  & 12.0 &19.4  & 14.0 &18.4  &32.1  &42.3  \\
$\omega(782) $             & 6.10 &13.0  & 10.8 &15.7  &27.0  &38.5  \\
$f_0(980)$                 & 0.56 &1.18 & 0.83 &1.27  &2.27  &3.26  \\ 
\hline
\end{tabular}
\end{table}

\begin{figure}[!t]
\centering
\includegraphics[width=9cm,height=13.5cm]{\pathnow 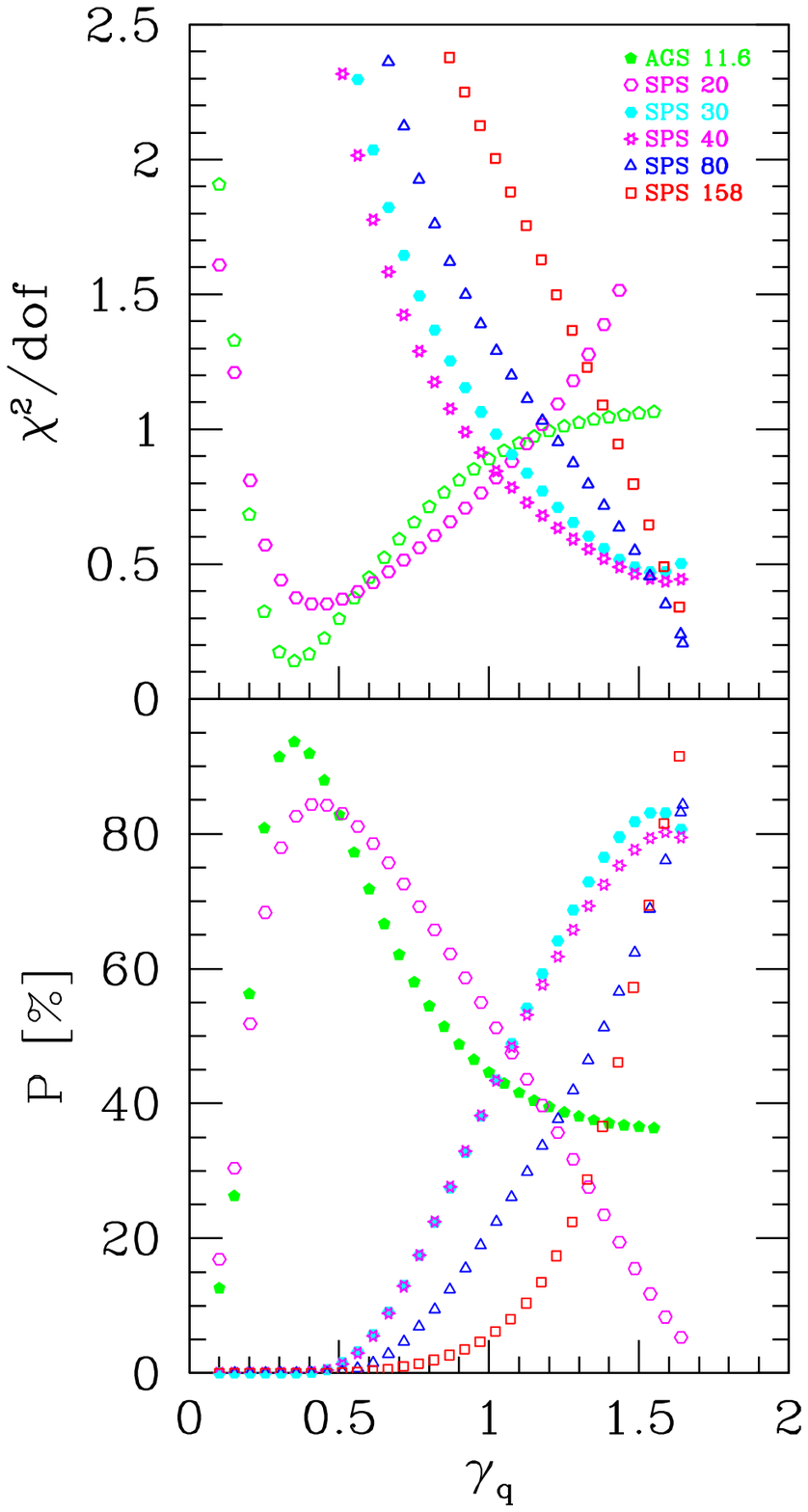}
\caption{\label{ChiP}
$\chi^2/{\rm dof}$ (top) and the associated  confidence level  $P[\%]$
(bottom) as function of $\gamma_q$, the light quark phase space occupancy.
for the AGS/SPS energy range.
}
\end{figure}
 
It is important to  inspect the  profiles of $\chi^2$, and of the confidence level $P[\%]$ 
determining the fit quality, see  \rf{ChiP}. 
We observe that the results for AGS 11.6 and SPS 20 GeV differ from the 
 remainder of the SPS results (30, 40, 80 and 158 GeV) in the outcome of the fit. The low
energy results, obtained at two different experimental  locations, clearly favor a 
value of $\gamma_q<1$, combined with relatively large $V,T$, while the higher energy
data favor  $\gamma_q\to \gamma_q^{\rm cr}$.   The  recently 
reported enlarged set of NA49 experimental results help
to discriminate the  chemical hadronization condition. Inspecting,  in particular, the 80 
and 158 GeV 
profiles, presented in \rf{ChiP}, we recognize that the semi-equilibrium model with 
$\gamma_q=1$  has a comparatively low 
viability compared to the full chemical non-equilibrium model we advance.

The SHARE package  offers the opportunity to evaluate the physical properties 
of the fireball in its local frame of reference: since we look at the hadron yields,
the flow velocity information is not retained. These results are shown in table \ref{AGSSPSPhysical}.
We note that the chemical freeze-out  at low energy  (AGS 11.6 and SPS 20 GeV) 
occurs from a much more dilute physical state,  the energy density of the high energy  (30, 40, 80 
and 158 GeV) 
data points hoovers well above 400--500 MeV/fm$^3$, about a factor 2.5 higher than at low energy.
We further note that between 20 and 30 GeV the ratio 
$E/TS$ shifts from a value below unity to above unity as required for
the sudden, supercooled hadronization mechanism for $E>20\ A$ GeV.
There is a steady growth in the yield of strangeness, both measured in terms of $s/S$ as well
as the yield per participant (net baryon number $b$). There is a decrease in the energy retained, indicating that 
the   flow effects grow rapidly, pushing the fraction of energy stopping below 50\% at the top
SPS energy. The cost of strangeness pair production $E_{\rm th}/\bar s$ decreases, as does 
the energy per hadron produced $E_{\rm th}/h$. Both these quantities use energy content in the  local rest frame,
and thus do not include the kinetic energy of  matter  flow at hadronization, which originated from 
the thermal pressure, which has driven the expansion matter flow.

\begin{table*}
\centering
\caption{
\label{AGSSPSPhysical}
The physical properties. Top: pressure $P$, energy density $\epsilon=E_{\rm th}/V$, entropy density $S/V$,
  for AGS and CERN energy range at,  
(top line) projectile energy $E$ [GeV]; middle: dimensionless ratios of properties at fireball breakup,
 $E_{\rm th}/TS$; strangeness per entropy $s/S$, strangeness per baryon $s/b$; and bottom
the  fraction of initial collision energy in  thermal
degrees of freedom, $(2E_{\rm th}/b)/\sqrt{s_{\rm NN}}$, the  
energy cost to make strangeness pair $E_{\rm th}/\bar s$,
thermal energy per hadron at hadronization $E_{\rm th}/h$.
}\vspace*{0.2cm}
\begin{center}
\begin{tabular}{|c| c | c c c c c |  }
\hline
E[$A$GeV]                   & 11.6        & 20          & 30          & 40         & 80            & 158  \\
$\sqrt{s_{\rm NN}}$  [GeV]  &4.84         &6.26         &7.61         &8.76        &12.32          &17.27  \\
\hline\hline
$P{\rm [MeV/fm}^3]$         &21.9         &21.3         &58.4         &68.0         &82.3         & 76.9            \\
$\epsilon{\rm [MeV/fm}^3]$  &190.1        &166.3        &429.7        &480.2        &549.9        & 491.8            \\
$S/V{\rm [1/fm}^3]$         &1.25         &1.21         &2.74         &3.07         &3.54         & 3.26            \\
\hline
$E_{\rm th}/TS$             &0.96         &0.92         &1.27         &1.20         &1.14         & 1.11         \\
100$\bar s/S$               &0.788         &1.26     &1.94        &1.90         &2.16          & 2.22       \\
$\bar s/b$                  &0.095         &0.202    &0.289       &0.314        &0.459         & 0.60       \\
 \hline
$(2E_{\rm th}/b)/\sqrt{s_{\rm NN}}$&0.752  & 0.722   &0.612       &0.589        &0.536        & 0.472     \\
$E_{\rm th}/\bar s {\rm\ [GeV]}$   &19.25  &10.9     &8.08        &8.21         &7.19        &6.80  \\
$E_{\rm th}/h {\rm\ [GeV]}$   &1.33  &1.18      &0.866      &0.859         &0.827        & 0.766\\
   \hline
\end{tabular}\vspace*{0.1cm}
 \end{center}
\end{table*}

\section{Discussion of Results} 
At the top SPS energy, the value of $s/S=0.022$   implies for a  QGP source a $\gamma_s^Q\simeq 0.7$, which 
corresponds to $\gamma_s/\gamma_q\simeq 1$. This, in fact, is the reason why 
chemical equilibrium $\gamma_q=\gamma_s=1$ `marginally works' 
for this data set.  However, as function of energy we see a very spectacular preference for non-equilibrium,
of two different types. For two lowest reaction energies considered, we are below 
chemical equilibrium and for other, higher energies, with $\sqrt{s_{\rm NN}}>7.6$ GeV we see 
over saturation of chemical occupancies.

Since we fit data very well, we also describe precisely the  K$^+/\pi^+$ ratio  as we show in \rf{KPi}. The maximum of the 
ratio    K$^+/\pi^+$ occurs  for  $E= 30 \ A$ GeV where we find  $\gamma_i>1$.   An  anomaly associated
with the  horn is the large yield of $\Lambda$, and protons, see bottom section of table \ref{AGSPS}. We further 
note that the structure of the horn  shown by dashed    (semiequilibrium) and dotted (equilibrium) lines  is also 
reproduced  qualitatively, contrary to reports made by other 
groups. We have traced this  behavior to our relaxation of the strangeness conservation condition.
\begin{figure} 
\centering
\includegraphics[width=9cm]{\pathnow 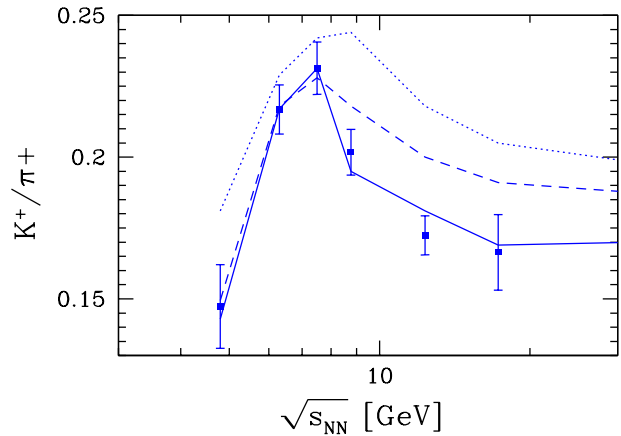}
\vskip -0.2cm
\caption{\label{KPi}
$K^+/\pi^+$ total yields as function of  $\sqrt{s_{\rm NN}}$.
 The solid lines show chemical non-equilibrium  model fit.
 The chemical equilibrium
fit result is shown by the dotted line. The dashed line arises
finding best $\gamma_s$ for $\gamma_q=1$. 
}
\end{figure}
 
Do the low energy results  imply  absence of quark matter,
and thus reactions between individual hadrons? Our analysis shows
that the chemical freeze-out occurs in a highly dilute phase. However, the rapid rise of strangeness
yield as function of reaction energy,   suggests that the strangeness production
processes differ from those encountered in normal hadron matter.   For this reason we 
favor a constituent quark matter reaction picture at  11.6 and 20 $A$ GeV, with 
color deconfinement arising yet below this energy range. The relatively 
high temperature and low $\gamma_q$ are consistent with properties of  
constituent  quark phase  with $m_{u,d}\simeq 340$ MeV and $m_s\simeq 500$ MeV, 
gluons are `frozen'. In such a massive deconfined quark phase  chiral symmetry is not restored.  
For  $\mu_{\rm B}\to 0$ the lattice results unite the chiral symmetry restoration, in which $m_q\to 0$, with 
the deconfinement transition.

In Summary: the physical properties we find for the hadronization of 30,\,40,\,80,\,158 $A$ GeV
most central heavy ion reactions  correspond to the expected behavior of the chirally symmetric QGP phase. 
SHM model described these results well,  hadron simulations (not discussed here) fail to account for multistrange
(anti)baryons. 

\subsubsection*{Acknowledgments}
Work supported by a grant from: the U.S. Department of Energy  DE-FG02-04ER4131.
LPTHE, Univ.\,Paris 6 et 7 is: Unit\'e mixte de Recherche du CNRS, UMR7589
.


\begin{thebibliography}{}


\bibitem{Letessier:2005qe}
  J.~Letessier and J.~Rafelski,
  arXiv:nucl-th/0504028.


\bibitem{Kuznetsova:2006bh}
  I.~Kuznetsova and J.~Rafelski,
  Eur.\ Phys.\ J.\  C {\bf 51} (2007) 113
  [arXiv:hep-ph/0607203].

\bibitem{NA49Gaz}
The results of NA49 we use were provided by M. Gazdzicki and  B. Lungwitz,
(private communication, September 2006) 


\bibitem{Torrieri:2004zz}
  G.~Torrieri et.al. 
  Comput.\ Phys.\ Commun.\  {\bf 167} (2005) 229
  [arXiv:nucl-th/0404083];\\
%
ibid {\bf 175} (2006) 635
  [arXiv:nucl-th/0603026].



\bibitem{Koch:1984tz}
  P.~Koch et.al. 
  Nucl.\ Phys.\  A {\bf 444} (1985) 678 
and   Phys.\ Rept.\  {\bf 142} (1986) 167.



\bibitem{Rafelski:1982ii}
  J.~Rafelski,
  Phys.\ Rept.\  {\bf 88} (1982) 331.

\bibitem{Letessier:2004cs}
  J.~Letessier, J.~Rafelski and G.~Torrieri,
  arXiv:nucl-th/0411047.






\end{thebibliography}
\end{document}